\documentstyle[12pt]{article}
\newcommand{\be}{\begin{equation}}
\newcommand{\ee}{\end{equation}}

\input{epsf.sty}
\def\framegraphics{\def\ifframe{\iftrue}}
\def\dontframegraphics{\def\ifframe{\iffalse}}
\def\drawgraphics{\def\ifdraw{\iftrue}}
\def\dontdrawgraphics{\def\ifdraw{\iffalse}}

\dontframegraphics
\drawgraphics

\newcommand{\graphics}[6]{
\def\epsfsize##1##2{#6##1}
\begin{picture}(#2,#3)
  \ifframe
    \put(0,0){\framebox(#2,#3){}}
  \fi
  \ifdraw
    \put(0,#3){\begin{picture}(0,0)
                 \put(#4,#5){\epsfbox{#1}}
               \end{picture}}
  \fi
\end{picture}}

\begin{document}

ADP-95-38/T192  \hspace*{6cm} September 1995

\vspace{2cm}
\centerline{\large The Role of Strange and Charm Quarks
in the}
\centerline{\large Nucleon Spin Structure Function}
\vspace{.4cm}
\centerline{F.M. Steffens and A.W. Thomas}
\vspace{1cm}
\centerline{Department of Physics and Mathematical Physics}
\centerline{University of Adelaide}
\centerline{Adelaide, S.A. 5005, Australia}
\vspace{.3cm}
\begin{abstract}
We perform an analysis of the relation between the factorization scale
and the masses of the quarks in the calculation of the hard gluon
coefficient in polarized deep inelastic scattering. Particular
attention is paid to the role of strange and charm quarks at finite
momentum transfer. It is found that for the momentum transfer of the
present experiments, the contribution from the charm quark is significant.

\end{abstract}

\section{Introduction}
In the usual analysis of the proton spin structure function,
based upon QCD and the operator product expansion (OPE), the moments of
the singlet part of $g_1 (x,Q^2)$ are written as:

\begin{equation}
\int_0^1 g_1^{(s)} (x, Q^2) x^{n-1} dx =
\Delta\Sigma_n (\mu^2) C^q_n (Q^2/\mu^2) +
\Delta g_n(\mu^2)C^g_n (Q^2/\mu^2),
\label{a2}
\end{equation}
with $C^q_n (Q^2/\mu^2) = 1 + O (\alpha_s)$ the Wilson
coefficients for the quark operators, $C^g_n (Q^2/\mu^2)=O(\alpha_s)$
the Wilson coefficients for the gluon operators. The matrix elements
$\Delta\Sigma_n (\mu^2)$ and $\Delta g_n(\mu^2)$ are not determined by
perturbative QCD and should be either fixed by experimental constraints
or calculated using non-perturbative techniques.
Eq. (\ref{a2}) can be inverted, using the inverse Mellin transformation,
and the result is:

\begin{equation}
g_1^{(s)} (x, Q^2) = \Delta\Sigma (x, \mu^2)\otimes C^{q} (x, Q^2/\mu^2) +
\Delta g(x, \mu^2)\otimes C^{g} (x, Q^2/\mu^2),
\label{a3}
\end{equation}
where $\otimes$ denotes a convolution of the two functions.

Much of the debate on the proton spin in the last few years has been
centered on whether or not the spin of the proton receives
a contribution from the gluons [1-6].
On the basis of the OPE the picture is clear: there is no twist two gluon
operator contributing to the first moment of $g_1$, and hence
$\int_0^1 C^g (x, Q^2/\mu^2) dx = 0$.
This result implies that the first moment of $g_1$ is given solely by
the first moment of $\Delta\Sigma$. If $\Delta\Sigma$ were identified
with the spin in the proton carried by the quarks then the gluons would
give no contribution. In this scenario, following the parton model language,
$\Delta\Sigma_n (\mu^2)=N \sum_{f} \Delta f_n (\mu^2)$, with
$\Delta f = \Delta f_1$ the amount of spin carried by the $f$ quark
and $N$ equals 1/9 for three flavors, 5/36 for four flavors, etc. It happens
that $\Delta\Sigma$ cannot be identified with
spin because of the axial anomaly. Indeed, the axial anomaly is at the
heart of the disagreement between the OPE
and the improved parton model (IPM) results for the role of gluons in
the first moment of $g_1 (x, Q^2)$. In this contribution, we are not going
to make a complete analysis of the equivalence (or otherwise) of these
approaches but will limit ourselves to
the analysis of the gluon contribution in the light of the IPM only.

In the IPM the situation is more complicated. One calculates
the full, polarized photon-proton cross section and uses the factorization
theorem to separate the hard and soft parts:

\begin{equation}
\sigma^{\gamma^v N} (x,Q^2) = \sigma^{\gamma^v q}_h (x,Q^2/\mu^2)\otimes
\Delta f_{q/N}(x,\mu^2) +
\sigma^{\gamma^v g}_h (x,Q^2/\mu^2)\otimes \Delta f_{g/N}(x,\mu^2),
\label{a1}
\end{equation}
where $\mu^2$ is the factorization scale, $\Delta f_{q(g)/N}$ is the
polarised quark (gluon) spin
distribution inside the nucleon and $\sigma_h$ is the polarized, hard
photon-quark
or hard photon-gluon cross section. One then could relate $g_1$ calculated in
the
IPM, Eq. (\ref{a1}), to $g_1$ calculated in the OPE, Eq. (\ref{a3}), by
identifying
the hard, perturbatively calculated, Wilson
coefficients with the hard photon-quark and hard photon-gluon cross sections
and identifying the matrix element $\Delta\Sigma (x, \mu^2)$ ($\Delta g(x,
\mu^2)$)
with the factorized quark (gluon) distribution $\Delta f_{q(g)/N}$.
However, as already mentioned, $\Delta\Sigma (x, \mu^2)$ cannot be identified
with
the quark spin distribution. The relation between them is beyond the
scope of the present work. Instead, we will concentrate on the relation
between the Wilson gluon coefficient and the hard gluon cross section of the
IPM. Although there are excellent treatments of this subject
in the literature \cite{carlitz,rev,mank,anselmino}, we think that the present
contribution adds significantly to the understanding of the
behaviour of $g_1 (x, Q^2)$ at finite $Q^2$.

In section 2 we will develop the basis for the calculation of the
hard gluon coefficient in the IPM. The resulting expression
interpolates the known limits of $-\frac{\alpha_s}{2\pi}N_f$ for $m_q^2 <<
\mu^2$
and 0 for $m_q^2 >> \mu^2$ and overcomes convergence
problems found in an early work \cite{schafer}.
The effects of this generalized, hard gluon coefficient are discussed in
section 3.
In particular, its effect on the contribution to $g_1^p (x, Q^2)$ from up,
down,
strange and charm quarks is studied. Our results indicate
that these corrections are sizable and must therefore be taken
into account when extracting the polarized gluon distribution from the proton.
We also point out
in this section how this anomalous contribution is affected by finite $Q^2$.
In section 4 we calculate the amount of polarized gluon in the proton necessary
to explain the available data. We compare our result with other estimates
made using simply the limiting cases for the hard gluon coefficients.
Section 5 is used to study the region in $x$ where this contribution is
located.
In section 6 we summarise the results obtained in this article.

\section{Theoretical Construction}

The hard gluon cross section is extracted from
the full photon-gluon fusion cross section, $\sigma^{\gamma^v g}$ and is
calculated through the box graphs which
start at order $\alpha_s$. The other contribution from which it must be
separated is
the quark  distribution inside the gluon \cite{mank}. Mathematically this is
expressed as:

\begin{equation}
\sigma^{\gamma^v g}(x, Q^2) = \sigma^{\gamma^v g}_h (x, Q^2/\mu^2) +
\Delta q^{g}(x, \mu^2),
\label{a4}
\end{equation}
where $\Delta q^{g}$ is the polarized quark distribution inside a
gluon and $\sigma^{\gamma^v g}_h$ is the hard photon-gluon cross section
defined, in the IPM, as the contribution coming from quarks in the box graph
with transverse momenta greater than the factorization scale.

The full photon-gluon cross section has been calculated to be
\cite{bass,vogelsan}:

\begin{eqnarray}
\sigma^{\gamma^v g}(x, Q^2) &&=\; -\frac{\alpha_s}{2\pi}N_f
\frac{\sqrt{1 - \frac{4m_q^2}{W^2}}}
{1 - \frac{4x^2 P^2}{Q^2}}\left[(2x -1)(1 - \frac{2xP^2}{Q^2})\right. \nonumber
\\*
   && \left. \left( 1 - \frac{1}{\sqrt{1 - \frac{4m_q^2}{W^2}}
\sqrt{1 - \frac{4x^2 P^2}{Q^2}}}
ln\left(\frac{1 + \sqrt{1 - \frac{4m_q^2}{W^2}} \sqrt{1 - \frac{4x^2
P^2}{Q^2}}}
{1 - \sqrt{1 - \frac{4m_q^2}{W^2}} \sqrt{1 - \frac{4x^2
P^2}{Q^2}}}\right)\right)
\right. \nonumber \\*
   && \left. + \left( x - 1 + \frac{xP^2}{Q^2}\right)
\frac{2 m_q^2 (1 - \frac{4 x^2 P^2}{Q^2}) - P^2 x(2x - 1)(1 -
\frac{2xP^2}{Q^2})}
{m_q^2 (1 - \frac{4x^2 P^2}{Q^2}) - P^2 x(x - 1 + \frac{x P^2}{Q^2})} \right] ,
\label{a41}
\end{eqnarray}
with $P^2 = -p^2$ the gluon virtuality, $m_q$ the quark mass and
$W^2 = \frac{Q^2 (1 - x) - P^2 x}{x}$ the invariant mass squared
of the photon-gluon system. For very large momentum transfer,
$Q^2 >> m_q^2 , P^2$, the full cross section reduces to:

\begin{eqnarray}
\sigma^{\gamma^v g} (x, Q^2 /\mu^2 ) &=& \frac{\alpha_s}{2\pi}N_f \left[(2 x -
1)
\left( ln\frac{Q^2}{m_q^2 + P^2 x(1 - x)} + ln\frac{1 - x}{x} - 1 \right)
\right. \nonumber \\*
     && \left.+ (1 - x)\frac{2m_q^2 - P^2 x (2x - 1)}{m_q^2 + P^2 x(1 -
x)}\right].
\label{a42}
\end{eqnarray}
It remains to calculate $\Delta q^g$. This is given by computing
the triangle diagram or, equivalently, the integral over the
transverse momentum of the square
of the norm of the light-cone $q\overline{q}$ wave function of the
gluon \cite{carlitz,mank,brodsky}. As $\Delta q^g$ is a soft contribution,
the integral over the transverse momentum has to have a cut off:

\begin{eqnarray}
\Delta q^g (x,\mu^2) &=& \frac{\alpha_s}{2\pi}N_f
\int_0^{\mu^2} dk^2_\perp \frac{m_q^2 + (2x - 1)k^2_\perp }{[m_q^2 + P^2 x(1-x)
+
k^2_\perp ]^2} \nonumber \\*
  &=& \frac{\alpha_s}{2\pi} N_f \left\{(2x-1)ln\left( \frac{\mu^2 + P^2 x(1-x)
+ m_q^2}
{m_q^2 + P^2 x(1-x)}\right) \right. \nonumber \\*
  &+& \left. (1-x)\frac{2m_q^2 + P^2 x(1-2x)}{m_q^2 + P^2 x(1-x)}
\frac{\mu^2}{\mu^2 + P^2 x(1-x) + m_q^2} \right\}.
\label{a43}
\end{eqnarray}
Equation (\ref{a43}) is a generalization of previous results
\cite{carlitz,mank} including the dependence on the factorization scale for
any values of the quark masses and gluon virtuality. Its first moment
is zero for $\mu^2 << m_q^2, P^2$. If $\mu^2 >> m_q^2, P^2$ the
first moment of $\Delta q^g (x, \mu^2)$ is 0 for $P^2 >> m_q^2$, while
it is $\frac{\alpha_s}{2\pi}N_f$ for
$m_q^2 >> P^2$. Using Eqs. (\ref{a4}), (\ref{a42}) and (\ref{a43})
we can calculate the hard gluon coefficient:

\begin{eqnarray}
  \sigma^{\gamma^v g}_h (x, Q^2, \mu^2) &=& \frac{\alpha_s}{2\pi} N_f \left\{
(2x-1)\left[ ln\left( \frac{Q^2}{\mu^2 + P^2 x(1-x) + m_q^2}
\right) + ln\left(\frac{1-x}{x}\right) - 1\right] \right. \nonumber \\*
  &+& \left. (1-x)\frac{2m_q^2 + P^2 x(1-2x)}{\mu^2 + m_q^2 + P^2 x(1-x)}
\right\} .
\label{a431}
\end{eqnarray}
Notice that the first moment of Eq. (\ref{a431}) does not depend on the ratio
$m_q^2/P^2$ in the region $\mu^2 >> m_q^2, P^2$ - it is a legitimate
hard contribution.
Equation (\ref{a431}) is also a generalization of previous results and
from its limit, $\mu^2 >> m_q^2, P^2$, it may be argued \cite{ross} that the
gluons
contribute to the first moment of $g_1 (x, Q^2)$  because
$\int_0^1 \sigma^{\gamma^v g}_h (x, Q^2) dx = -\frac{\alpha_s}{2\pi}N_f$.

On the other hand, if one calculates the quark distribution
inside a gluon through the triangle graph, which we call $\Delta q^{g}_{OPE}$,
using a regularization scheme that respects the axial anomaly, it is found that
\footnote{The triangle graph regularized
with a cut off on the transverse momentum results in Eq. (\ref{a43}).}:

\begin{eqnarray}
\Delta q^g (x) - \Delta q^{g}_{OPE} (x) &=& \frac{\alpha_s}{\pi} N_f \left[
(2x-1)ln\left(\frac{\mu^2 + P^2 x(1-x) + m_q^2}{\mu^2}\right) \right. \nonumber
\\*
    && \left. + \frac{2 \mu^2 (1 - x)}{\mu^2 + P^2 x(1-x) + m_q^2} \right],
\label{a44}
\end{eqnarray}
where the renormalization scale in the
regularization of $\Delta q^{g}_{OPE}$ (using $\overline{MS}$)
has been taken to coincide with the factorization scale in the IPM.

Equipped with Eq. (\ref{a44}) we can understand exactly why
the hard gluon coefficient in the IPM has a first moment different
from zero. The reason is that in the process of factorization
the axial anomaly was shifted from the quark distribution
inside the gluon to the hard coefficient. Equation (\ref{a44}) reflects
the fact that the regularization of $\Delta q^{g}_{OPE}$ respects
the axial anomaly while the regularization of $\Delta q^{g}$ does not.
We also see that in the limit $m_q^2 >> \mu^2$, the
discrepancy between the two calculations disappears (at least for the
first moment - the x dependence depends on the regularization method).
A similar phenomenon is found in unpolarized deep inelastic
scattering
where an analysis by Bass \cite{bass3} has shown that the trace anomaly
induces the same sort of shift when a cut off over the transverse
squared momenta of the quarks is used to separate the soft and hard regions.

As a consistency check of our equations, we calculate the OPE hard coefficient,
$C^g$. It is defined
in the same way as $\sigma^{\gamma^v g}_h$ in Eq. (\ref{a4}) and calculated
with the help of expressions (\ref{a42}) and (\ref{a44}):

\begin{equation}
C^g (x, Q^2 /\mu^2)= \frac{\alpha_s}{2\pi}N_f \left[
(2x-1)\left( ln \frac{Q^2}{\mu^2}
+ ln\left(\frac{1-x}{x}\right) - 1 \right) + 2(1-x) \right] .
\label{a46}
\end{equation}
This result is independent of mass and its first moment is always
zero, in accordance with the results of Kodaira \cite{kodaira} and
Bodwin and Qiu \cite{bodwin}.

\section{Consequences for the First Moment of the Hard Gluon Coefficient}

It is interesting to study the dependence on $\mu^2$ of the first
moment of $\sigma^{\gamma^v g}_h$.
In an early study on this subject\footnote{We thank S. Bass for
pointing out to us this work.}, Mankiewicz and Sch\"{a}fer \cite{schafer}
determined the first moment of the box graph as a function of the minimum
transverse momentum carried by the quarks. Their results for
$Q^2\rightarrow\infty$ agree qualitatively with ours, as will soon be seen.
But it was also found in Ref. \cite{schafer} that for momentum transfers of
the order of 10 to $100\; GeV^2$,
the contribution from light quarks\footnote{We assume for the quark
masses their current values. We do not take into account variation of
the masses with the factorization scale but note that our conclusions
are not significantly altered by small changes in the quark mass.}
($m_q = 10\;MeV$) is deeply affected
by the choice of the minimum value for the transverse quark momentum.
In the method used here, such an ambiguity does not
exist for the light quarks and its anomalous contribution for $Q^2= 10$
or $100 \; GeV^2$ is well defined and independent of $k_\perp$.
We use this result to argue that the hard gluon coefficient calculated
here is more stable from the point of view of infrared singularities.

Even with the known variations of the anomalous contribution
with the factorization scale, it has been widely assumed in the
literature \cite{anselmino} that for light quarks ($u$, $d$ and $s$) the
first moment of $\sigma^{\gamma^v g}_h$ is $-\alpha_s / 2\pi$
and for heavy quarks (like $c$ or $b$) it is zero (because,
for $m_q^2 >> \mu^2$, $\sigma^{\gamma^v g}_h$ reduces to $C^g$).
But it also happens that the gluonic contribution to $g_1 (x,Q^2)$
is of the form $\sigma^{\gamma^v g}_h (x,Q^2/\mu^2)\otimes \Delta g (x,\mu^2)$.
This means that the scale $\mu^2$ at which the gluon distribution
is calculated (or parametrized) is the same scale $\mu^2$ that has
to be used in the calculation of the hard gluon coefficient, and that it
does not make sense to talk about the magnitude of the hard
gluon coefficient without specifying the factorization scale.
Thus, the heavy quark contribution is negligible only when the polarized gluon
contribution is calculated at a very low scale compared with the quark mass.

\begin{figure}[h]
\graphics{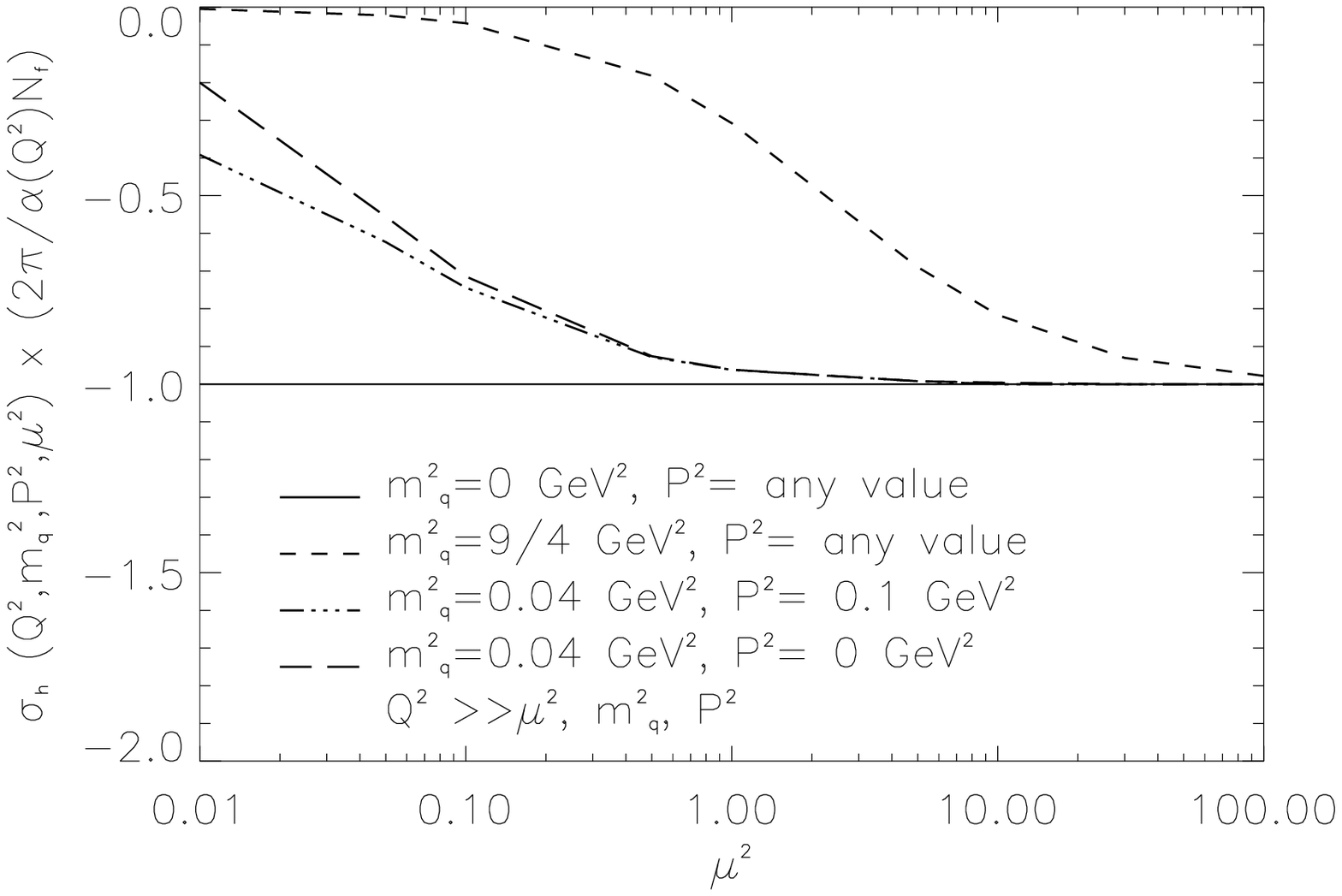}{17}{13}{-2}{-13}{0.9}
\caption{Hard gluon coefficient as given by Eq. (\protect\ref{a431}),
calculated with the assumption of infinite momentum transfer as a function
of the factorization scale. For realistic scales ($\mu^2 > 1\; GeV^2$),
the charm contribution is seen to be important.}
\label{fig1}
\end{figure}

In Fig. \ref{fig1} we show the first moment of $\sigma^{\gamma^v g}_h$ as a
function of the factorization scale for the $u$ and $d$ quarks
($m_q^2 \sim 25\times 10^{-6}\; GeV^2$),
for the $s$ quark ($m_q^2 \sim 0.04 \; GeV^2$) and for the $c$ quark
($m_q^2 \sim 9/4\; GeV^2$).
We see that, as is well known \cite{carlitz,rev,mank,anselmino},
the $c$ quark does not contribute when $m_c^2 >> \mu^2$,
as one can also verify directly from Eq. (\ref{a431}).
However, for reasonable values of $\mu^2$ there is
a contribution large enough to be taken into account. Thus, the
significance of the charm contribution to $g_1 (x, Q^2)$ depends
on where the polarized gluon distribution is calculated. For instance,
calculations have been made in the literature
using input polarized gluon distributions at a scale of typically $ 4 \;
GeV^2$.
The authors of these calculations usually disregard the charm contribution.
We note in passing that in the region of $\mu^2$
where polarized charm can be disregarded, the
polarized strange quarks yield only half of the contribution given by
$u$ and $d$ quarks. As we see from Fig. \ref{fig1}, the
$c$ quark gives around $64\%$ of the contribution of the light quarks for
$\mu^2 \sim 4\;GeV^2$
and so it should not be disregarded if the gluon distribution is
calculated at this scale. We also see from Fig. \ref{fig1} that
the $u$ and $d$ quarks give the same contribution, independent of the
factorization scale. We further notice that, for practical purposes, the hard
gluon coefficient is independent of the exact value of the gluon virtuality
$P^2$.

The discussion of the preceeding paragraph was based on the not
so realistic assumption that the momentum transfer $Q^2$ is infinitely
bigger than any other scales in the theory. It implies, for instance,
that when integrating the hard cross section we allow $x$ to go
from zero to one. But from simple kinematic arguments we know that $x$ has
a maximum value of $x_{max} = Q^2 / (Q^2 + P^2 + 4m_q^2)$ and so
$x_{max}\rightarrow 1$ only when $Q^2 >> m_q^2, P^2$. For the finite
$Q^2$ of the current experiments, $x_{max}$ never reaches 1 and so the
integral in $x$ has a cut off. For instance if one calculates the first
moment of $\sigma^{\gamma^v g}_h$ for the $c$ quark ($m_q^2 = 9/4 \; GeV^2$)
at $\mu^2=Q^2=4\;GeV^2$, using Eq. (\ref{a431}), one finds that its value
changes from -0.64, when $x$ is artificially allowed to reach 1, to
$\sim$ 0.015 when the physical
cut off in $x$ is applied. What happens is that expression (\ref{a431}) itself
was obtained under the assumption of an infinitely large $Q^2$. To
be more consistent when dealing with finite $Q^2$, one should derive
the hard cross section from the full cross section without any approximation.

\begin{figure}[h]
\graphics{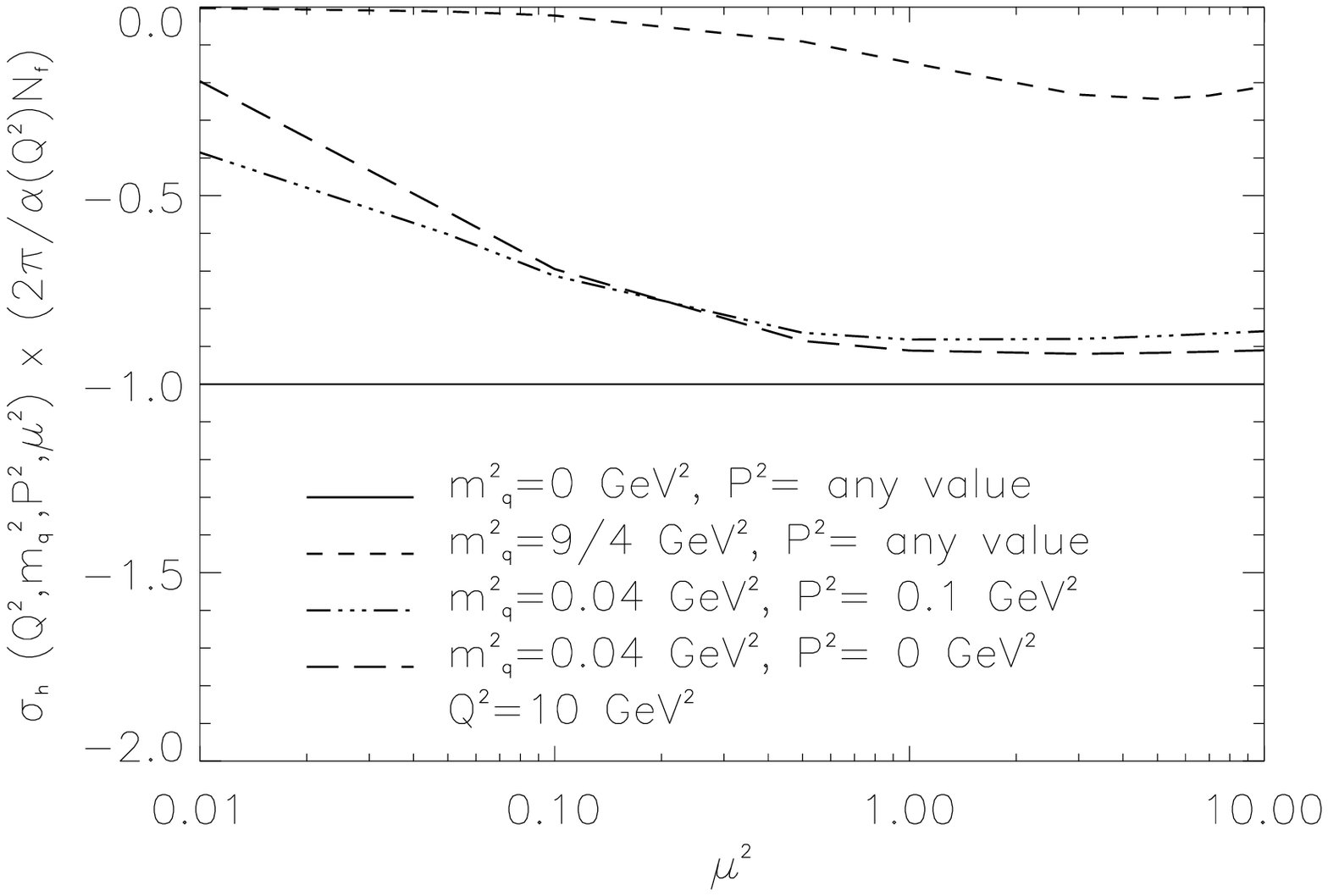}{17}{13}{-2}{-13}{0.9}
\caption{Hard gluon coefficient as given by Eqs. (\protect\ref{a41}),
(\protect\ref{a43}) and (\protect\ref{a48}).
The momentum transfer is fixed at $10 \; GeV^2$. It is seen that the
strange quark contribution never equals that from the up and down quarks
and the charm quark contribution is sizable for $\mu^2 > 1\; GeV^2$.}
\label{fig2}
\end{figure}

\begin{figure}[h]
\graphics{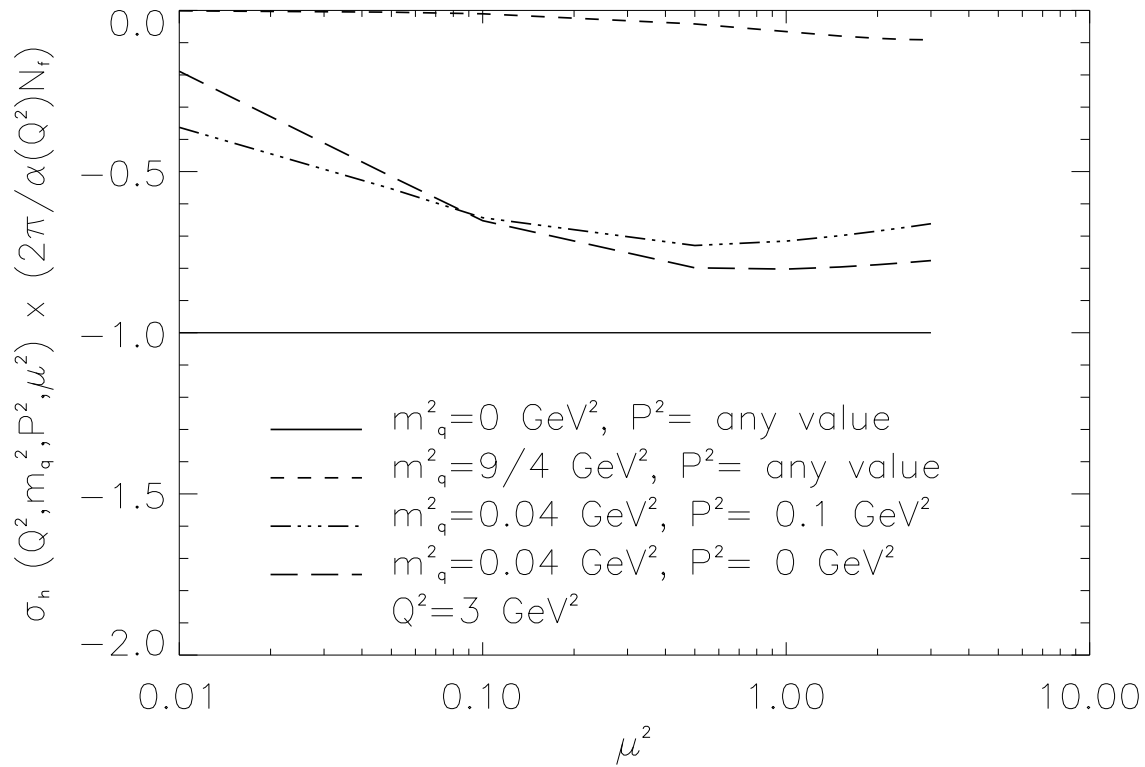}{17}{13}{-2}{-13}{0.9}
\caption{Hard gluon coefficient as given by Eqs. (\protect\ref{a41}),
(\protect\ref{a43}) and (\protect\ref{a48}).
The momentum transfer is fixed at $3 \; GeV^2$. It is seen that the
strange quark contributes with approximately $75\%$ of the up and down quarks
in the realistic region of $\mu^2 > 1\; GeV^2$. In the same region, the
charm quark gives a $10\%$ contribution.}
\label{fig3}
\end{figure}

\begin{figure}[h]
\graphics{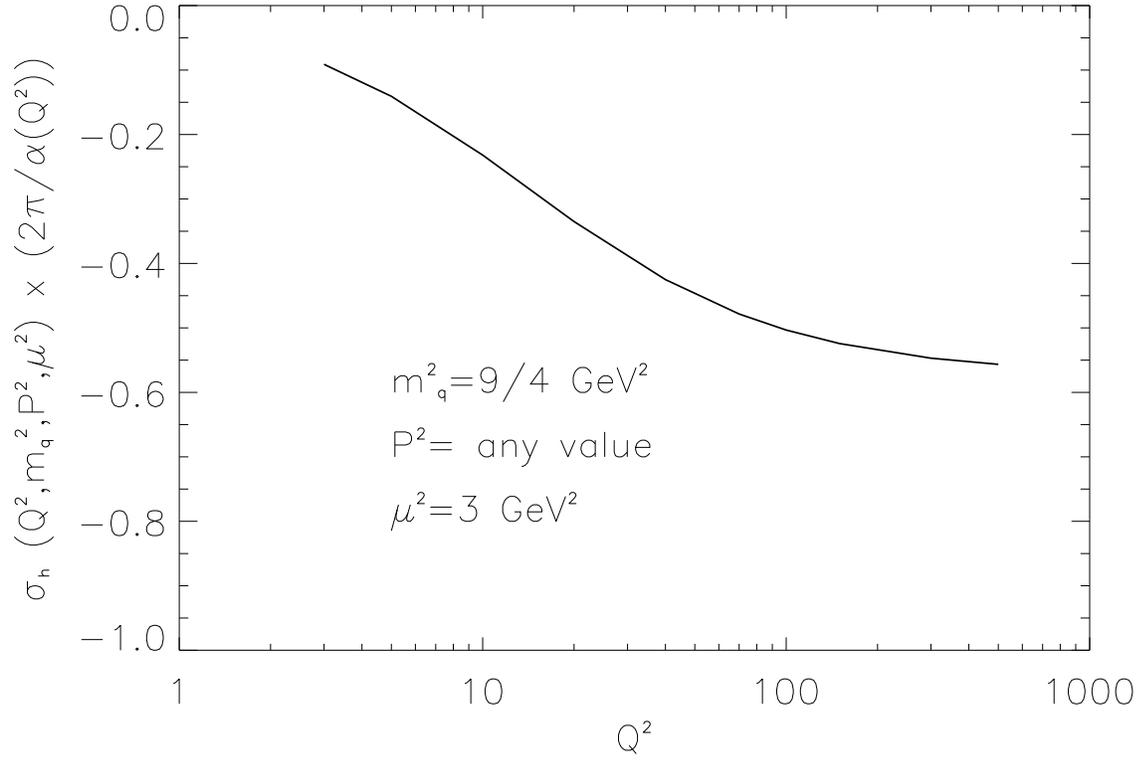}{17}{10}{-2}{-8}{0.9}
\caption{Hard gluon coefficient for the charm quark, calculated with
Eqs. (\protect\ref{a41}), (\protect\ref{a43}) and (\protect\ref{a48}).
The factorization scale is fixed at $3\; GeV^2$ and the $Q^2$ dependence
is studied.}
\label{fig4}
\end{figure}

In the general case we then write:

\begin{equation}
C^g = \sigma^{\gamma_v g} - \Delta q^g_{OPE},
\label{a47}
\end{equation}
\begin{equation}
\sigma^{\gamma^v g}_h = \sigma^{\gamma_v g} - \Delta q^g ,
\label{a48}
\end{equation}
with $\sigma^{\gamma_v g}$ given by Eq. (\ref{a41}) and $\Delta q^g$
and $\Delta q^g_{OPE}$ given by Eqs. (\ref{a43}) and (\ref{a44}). We stress
that these equations are the complete result at order $\alpha_s$.
In Figs. \ref{fig2} and \ref{fig3} we show the first moment of
$\sigma^{\gamma^v g}_h$, defined in Eq. (\ref{a48}), as a function of
the factorization scale $\mu^2$ for $Q^2 = 10\; GeV^2$ and $Q^2 = 3\; GeV^2$,
respectively. These values were chosen because they are the average $Q^2$ of
the EMC \cite{emc,smc} and SLAC \cite{slac} experiments. The resulting
dependence is very interesting. It shows that in the region of interest
($\mu^2 \geq 1\;GeV^2$) there is no appreciable dependence on the gluon
virtuality or on $\mu^2$ (at least for $Q^2 = 10\;GeV^2$) but the mass
dependence is strong. Remarkably, the contribution from the $s$ quark is
never the same as the contribution from the $u$ and $d$ quarks, contrary
to what is usually claimed.
The $s$ contribution is $\sim 0.9 \; \alpha_s (Q^2) / 2\pi$ for the EMC
data and $\sim 0.75 \; \alpha_s (Q^2) / 2\pi$ for the E143-SLAC data.
We also find a nonnegligible contribution coming from the $c$ quarks.
For the EMC data, the $c$ quark contributes with
$\sim 0.2 \; \alpha_s (Q^2) / 2\pi$
and for the E143-SLAC data with  $\sim 0.1\; \alpha_s (Q^2) / 2\pi$.
Figure \ref{fig2} is unaltered \footnote{In reality, there is
a $\sim 1\%$ correction for $\mu^2\sim 0.01\;GeV^2$. This result is
in complete accord with the fact that the anomalous contribution for the $u$
and $d$ quarks goes to zero as $\mu^2$ goes to zero.}
if we go from $m_q^2=0$ to $m_q^2= 1\times 10^{-4} GeV^2$.
If we then compare our Fig. \ref{fig2} with Fig. 2 of Ref. \cite{schafer} we
see clearly that the present approach does not have a convergence problem
in $Q^2$ and yelds a perfectly unambiguous contribution from the
light quarks.
Finally, we show in Fig. \ref{fig4} the $Q^2$ dependence of the polarized
charm contribution calculated with $\mu^2 = 3\; GeV^2$. This contribution,
obviously, tends to the value calculated in Fig. \ref{fig1}
($\sim -0.57\; \alpha_s (Q^2) / 2\pi$).

\section{Relevance in Analysing the Fraction of Nucleon Spin Carried by Gluons}

In terms of the polarized quark and gluon distributions, $g_1^p (x, Q^2)$
for 4 flavors is written as:

\begin{eqnarray}
g_1^p (x, Q^2)&=&\frac{1}{12}\Delta q_3 (x,Q^2) +
\frac{1}{36}\Delta q_8 (x,Q^2) - \frac{1}{36}\Delta q_{15} (x,Q^2) \nonumber
\\*
&& + \frac{5}{36}\Delta\Sigma (x,Q^2) + \frac{5}{36}
\sigma^{\gamma^v g}_h (x,Q^2,\mu^2) \otimes\Delta g (x, \mu^2),
\label{a5}
\end{eqnarray}
where $\Delta q_3=\Delta u - \Delta d$, $\Delta q_8 = \Delta u + \Delta d
- 2\Delta s$, $\Delta q_{15} = \Delta u + \Delta d + \Delta s - 3 \Delta c$
and $\Delta\Sigma = \Delta u + \Delta d + \Delta s + \Delta c$.
For 3 flavors the coefficient of the singlet part changes from 5/36 to
1/9 and $\Delta q_{15}$ does not exist. To order $\alpha_s (Q^2)$
\cite{kodaira}, the first moment of (\ref{a5}) is:

\begin{eqnarray}
\Gamma_1^p (Q^2) &=& I_3 (Q^2) + I_8 (Q^2) - I_{15} (Q^2) + I_0 (Q^2) \nonumber
\\*
&& - \frac{5}{36}\left( 2\frac{\alpha_s (Q^2)}{2\pi} + s_1
\frac{\alpha_s (Q^2)}{2\pi} + c_1 \frac{\alpha_s (Q^2)}{2\pi}\right)
\Delta g(Q^2).
\label{a6}
\end{eqnarray}
The coefficients of $\alpha_s (Q^2)$ have the following meaning. The 2
indicates that the $u$ and $d$ quarks give the same contribution
$\frac{\alpha_s (Q^2)}{2\pi}$, as discussed before. The $s_1$ and $c_1$
factors give the amount of strange and charm quark contributions,
according to Eq. (\ref{a48}) and Figs. 1-3.

To extract the value of $\Delta G(Q^2)$ we will closely follow
Refs. \cite{stirling2,close}. For the sake of comparision, we
begin with only 3 flavors and with the common
assumption that the $u$, $d$ and $s$ quarks give the same
anomalous contribution.

Under the assumption that the polarized sea originates
exclusively from the anomalous gluon contribution we have, for
3 flavors, the following identities:

\begin{eqnarray}
I_3 &=& \frac{1}{12} (F + D) (1 - \frac{\alpha_s}{\pi}) \nonumber \\*
I_8 + I_0 &=& \frac{1}{36} (3F - D)\left[(1 - \frac{\alpha_s}{\pi})
+ 4(1 - \frac{\alpha_s}{3\pi})\right],
\label{a7}
\end{eqnarray}
where the the quark spin fractions were expressed in terms of the
$F$ and $D$ couplings and corrections from the two loop expansion of
the beta function and anomalous dimension were incorporated.
In $NLO$, $\alpha_s$ is given as the solution of the
following transcendental equation:

\begin{equation}
ln\frac{Q^2}{\Lambda^2}=\frac{4\pi}{\beta_0 \alpha_s} -
\frac{\beta_1}{\beta_0^2}ln\left[\frac{4\pi}{\beta_0 \alpha_s}
+ \frac{\beta_1}{\beta_0^2}\right],
\label{a9}
\end{equation}
with $\beta_0 = 11-2N_f /3$ and $\beta_1 = 102 - 38N_f /3$.
We use $\Lambda = \Lambda^{(3)} = 248 \; MeV$, determined by
fixing $\Lambda^{(4)}=200 \; MeV$ \cite{grv}. Using the experimental
values of $F$ and $D$ as given in \cite{close}, we determine
$I_3$ and $I_8 + I_0$ at $Q^2=10 \; GeV^2$ (with $\alpha_s (Q^2 = 10\; GeV^2)
\simeq 0.209$):

\begin{eqnarray}
I_3 &=& 0.0977 \pm 0.001 \nonumber \\*
I_8 + I_0 &=& 0.0779 \pm 0.002
\label{a10}
\end{eqnarray}
We now use Eq. (\ref{a6}) to determine $\Delta G(Q^2)$. On the
left hand side, we use the experimental result \cite{smc}:

\begin{equation}
\Gamma_1^p (Q^2 = 10 GeV^2) = 0.142\pm 0.008\pm 0.011.
\label{a11}
\end{equation}
On the right hand side we use the results (\ref{a10}),
$s_1 = 1$, $c_1 =0$ and remember that for 3 flavors
the singlet coefficient is 1/9 and $I_{15}=0$. The result is:

\begin{equation}
\Delta g(Q^2 = 10 GeV^2) = 3.04 \pm 1.4.
\label{a12}
\end{equation}
For 4 flavors the analysis is similar. One just has to redefine
the integral of $g_1^p (x,Q^2)$:

\begin{eqnarray}
I_3 &=& \frac{1}{12} (F + D) (1 - \frac{\alpha_s}{\pi}) \nonumber \\*
I_8 + I_0 - I_{15} &=& \frac{5}{36} (3F - D)\left(1 -
\frac{\alpha_s}{3\pi}\right).
\label{a13}
\end{eqnarray}

We then proceed as before and calculate $\Delta G$ using
the result for the gluon coefficient as displayed in Fig.\ref{fig2}.
We see that for $\mu^2 = Q^2 = 10 \; GeV^2$, $s_1 \sim 0.9$,
$c_1 \simeq 0.21$ and $\alpha_s (Q^2 = 10 \; GeV^2) = 0.2142$,
resulting in:

\begin{eqnarray}
I_3 &=& 0.0976 \pm 0.001 \nonumber \\*
I_8 + I_0 - I_{15} &=& 0.0786 \pm 0.002 \nonumber \\*
\Delta g (Q^2 = 10\; GeV^2) &=& 2.32 \pm 1.06
\label{a15}
\end{eqnarray}
In passing we notice that if the usual assumption of
infinite momentum transfer were used, then according to the
results of Fig.\ref{fig1}, at $10 \; GeV^2$ $s_1 = 1$,
$c_1 \simeq 0.81$ and hence $\Delta g \simeq 1.89$.

\section{The $x$ dependence}

\begin{figure}[h]
\graphics{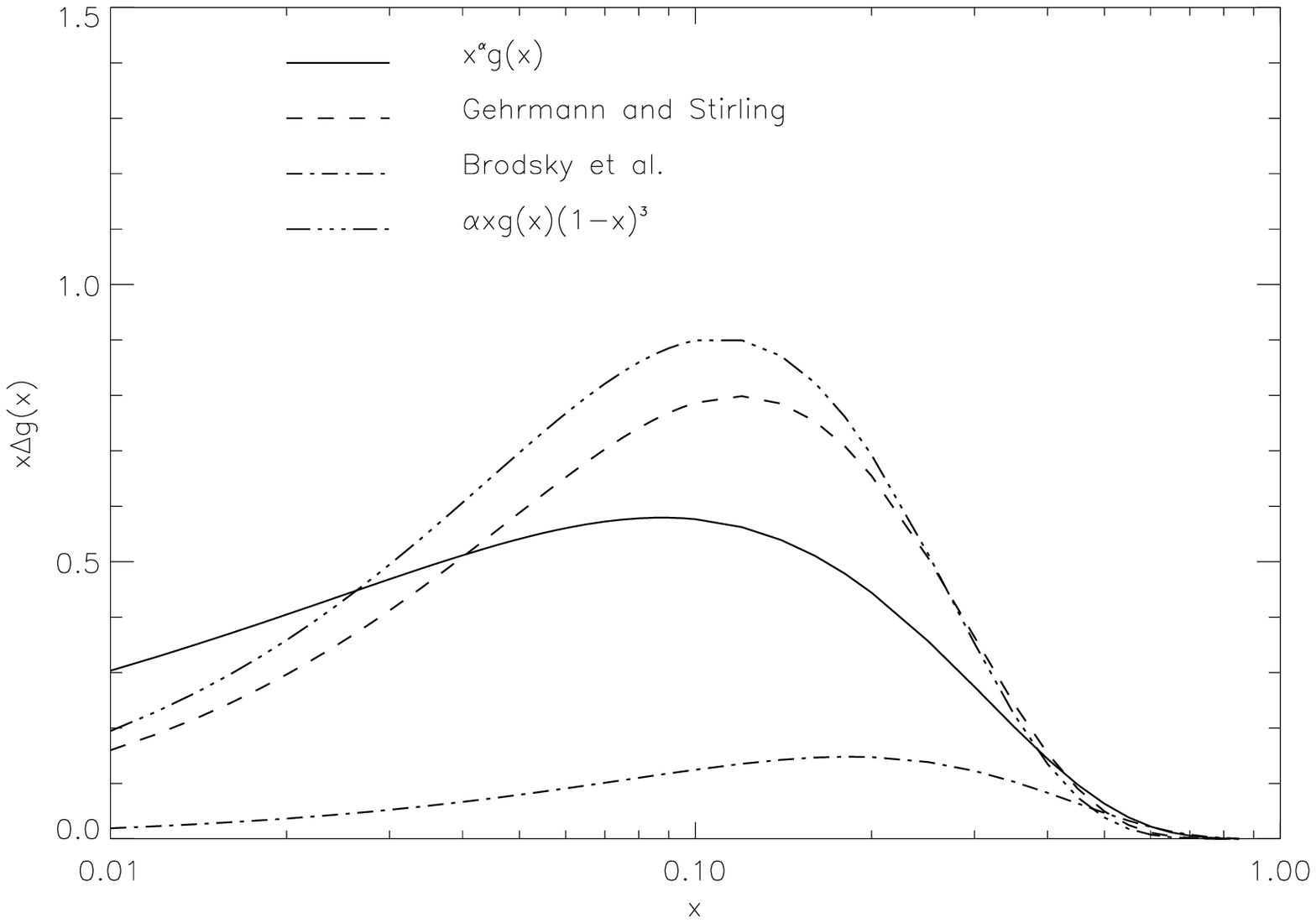}{17}{10}{-2}{-8}{0.9}
\caption{Comparison of various polarized gluon distributions
considered in the text.}
\label{fig5}
\end{figure}

The exact $x$ dependence of the anomalous contribution
is a matter of convention because the freedom in the factorization
scheme while calculating $\sigma_h^{\gamma_v g}$. Other choices of
regularization would result in different functions of $x$. But, as
shown by Gl\"{u}ck et al. \cite{stirling}, the exact form of
the $x$ dependence seems not to be very important.
Once we do not know the form of the polarized gluon distribution,
the best we can do is constrain it by some general considerations.
For instance, there is the positivity condition:

\be
|\Delta g (x, Q^2)|\leq g(x,Q^2),
\label{a16}
\ee
where $g(x,Q^2)$ is the unpolarized gluon distribution.
A very simple form that satisfies the above condition is:

\be
\Delta (x) = x^{\alpha} g(x),
\label{a17}
\ee
where $\alpha$ is determined through the normalization of
$\Delta g$. For $\Delta g$ of Eq. (\ref{a15}), $\alpha = 0.49$.
The advantage of using this form to study the
$x$ dependence is its simplicity. The problem with
Eq. (\ref{a17}) is that it does not have the correct behavior as
$x\rightarrow 0$. As proposed by Brodsky et al. \cite{brodsky2},

\be
\frac{\Delta g(x)}{g(x)}\rightarrow x,
\label{a18}
\ee
as $x\rightarrow 0$. From the many ways
to satisfy both conditions (\ref{a16}) and (\ref{a18}), we choose:

\be
\Delta g (x,\mu^2=9\;GeV^2) = \alpha x g(x,\mu^2=9\;GeV^2) (1 - x)^3 ,
\label{a19}
\ee
where $\alpha=6.92$ for $\Delta g=2.32$.
We made this choice guided only by the desire of simplicity and to produce
a polarized gluon distribution that resembles an already existing one
\cite{stirling2}. For the unpolarized gluon distribution, we use the
one given by the NMC \cite{nmc}, determined from inelastic $J/\psi$ production:

\be
x g(x) = \frac{1}{2}(\eta + 1)(1 - x)^\eta.
\label{a20}
\ee
This parametrization is valid for $\mu^2 = 9\; GeV^2$ and should not
be trusted for $x\leq0.01$. Again, this choice is based on simplicity
and we note that a further parametrization by the NMC \cite{nmc2} group
agrees with Eq. (\ref{a20}) for $x\geq 0.01$. The parameter $\eta$ is
$\eta=5.1\pm 0.9$. Given these choices, we show in Fig. \ref{fig5} the
forms (\ref{a17}) and (\ref{a19}) for the polarized gluon distributions
plus the forms of Brodsky et al. \cite{brodsky2} and Gehrmann and
Stirling (GS) \cite{stirling2}, calculated at $4\;GeV^2$. Our parametrization
(\ref{a19}) is slightly higher than that of GS because of the normalization
factor. If we use the
same normalization as theirs\footnote{We note that in \cite{stirling2},
the coupling constant is calculated in leading order rather than
in NLO. This would lead to
an increase of the total polarization carried by the gluons.}, both curves
would be essentially the same. Evolution from 4 to 9 $GeV^2$ for the GS
distribution also has small effects. The parametrization of Brodsky et al.
\cite{brodsky2} is much smaller than the others because in their approach
the polarized gluons are not responsible for the small experimental
value of Eq. (\ref{a11}).

\begin{figure}[h]
\graphics{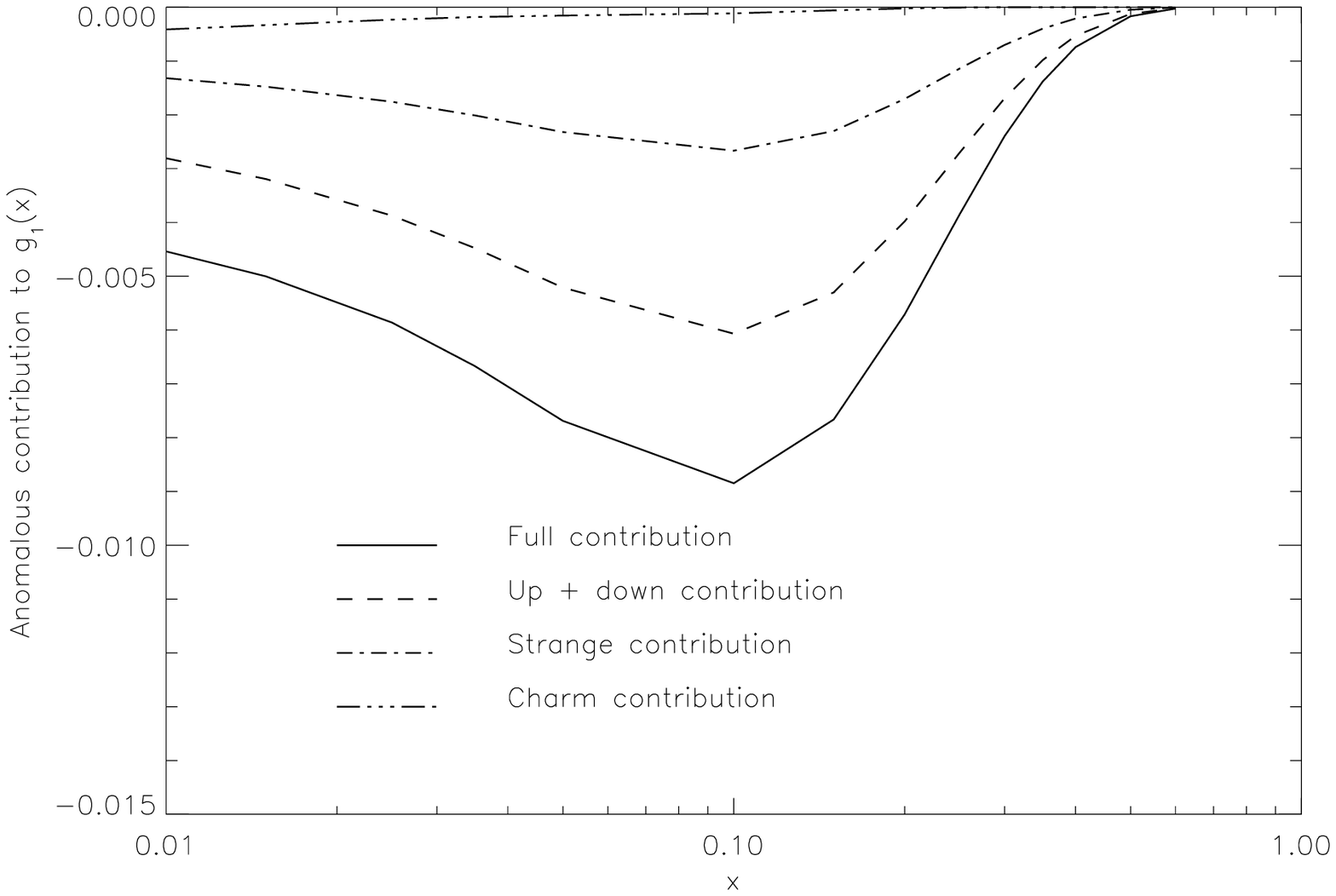}{17}{10}{-2}{-8}{0.9}
\caption{Comparison of the $x$-dependence of the non-strange,
strange and charm quark distributions to $g_1 (x)$. The anomalous contribution
is given by $\frac{5}{36}
\sigma^{\gamma^v g}_h (x,Q^2,\mu^2)\otimes\Delta g (x, \mu^2)$ and it
is used the form $\Delta g (x,\mu^2=9\; GeV^2) = \alpha x g(x) (1 - x)^3$ for
the
polarized gluon.}
\label{fig6}
\end{figure}

Using the constructed gluon distributions, we can estimate where
in $x$ the anomalous contribution is located. In Fig. \ref{fig6}
we show the anomalous contribution, $\frac{5}{36}
\sigma^{\gamma^v g}_h (x,Q^2,\mu^2)\otimes\Delta g (x, \mu^2)$, for
$\Delta g (x) = \alpha x g(x) (1 - x)^3$. We see that its contribution
inside the experimental region is important.To better
evaluate its importance, we calculated the amount of the total gluon that
lies inside the region $0.01\leq x \leq 1$. For 4 flavors, the contribution
from $x\geq 0.01$ corresponds to about $66\%$ of the total anomalous
contribution. In the case of 3 flavors, this percentage is $\sim 69\%$.
For $\Delta g(x)=x^{\alpha} g(x)$ this conclusion is not dramatically altered.

\begin{figure}[h]
\graphics{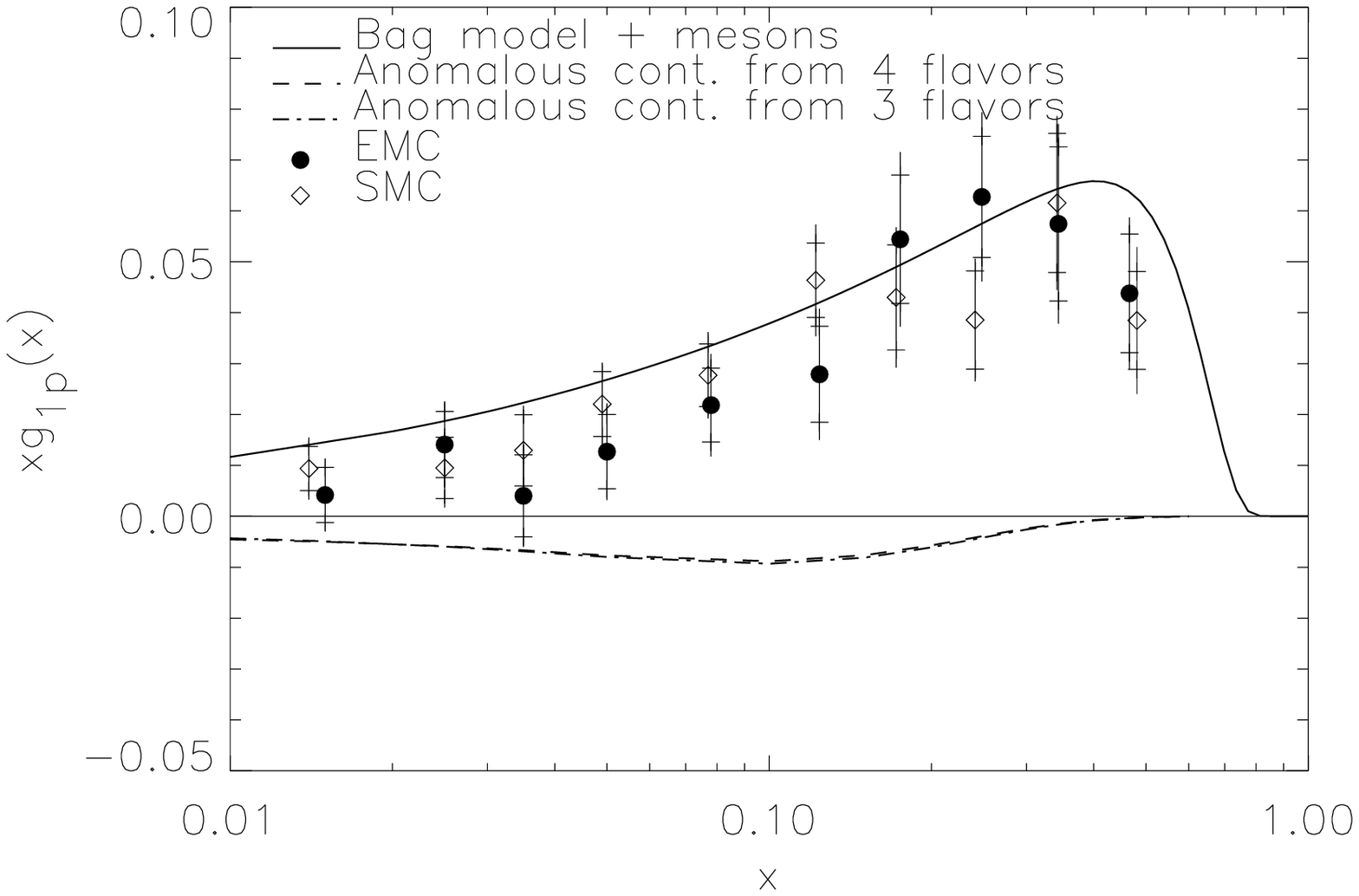}{17}{10}{-2}{-8}{0.9}
\caption{EMC \protect\cite{emc} and SMC \protect\cite{smc}
data for $g_{1p}(x)$ at $10\; GeV^2$. The theoretical
curve for the polarized valence distribution is calculated in NLO and
taken from Ref.\protect\cite{me}. The anomalous contribution should be
subtracted is to be subtracted from the theoretical curve.}
\label{fig8}
\end{figure}

It is also interesting to compare the anomalous gluon contribution
directly with the experimental data. To this end,
we plot in Fig. \ref{fig8} the experimental data \cite{emc,smc}
for $g_1^p (x)$ together with an early next-to-leading order estimate
\cite{me} for the valence quark distribution and the anomalous gluon
contribution for the case of 3 and 4 flavors. A remark is necessary here.
The calculation of the hard gluon coefficient was performed through a
cut off on the transverse momenta of the partons in order to regularize
the integrals. This procedure is the definition of the parton model.
On the other hand, we calculated the strong coupling constant, and also
the evolution of $g_1^p$ in Ref. \cite{me}, using the $\overline{MS}$
scheme. In principle, showing the $x$ dependence of two quantities
in different schemes is not a consistent procedure. However, the problem
is not as bad as it looks. First, if we change schemes we can mantain
$\alpha_s$ unaltered by a simple redefinition of the parameter
$\Lambda$ in  expression (\ref{a9}). Second, the theoretical curve
for $g_1^p (x, Q^2)$ is to be interpreted as
a guide of what a parametrization for the valence part of the polarized
structure function would give, the regions in which it differs from
the data and where it should be corrected. A proper procedure would be
to calculate the quark distribution, the anomalous dimensions and the Wilson
coefficients in the same
scheme. That said we proceed noticing that the integral
over $x$ of the valence contribution calculated in \cite{me} ($\sim0.169$)
is in complete agreement with the estimates calculated previously in
Section 4. The two curves below the
origin, are the anomalous contributions that
should be added to the solid curve for $N_f = $3 or 4. As we fixed the
normalization of the total polarized glue for either 3 or 4 flavors,
there is no noticeable difference between the two cases.
We see that the anomalous contribution is potentially important
to correct the $x$ dependence of the polarized valence distribution
inside the proton.

\section{Discussion}

In summary, there is a gluonic
contribution to the proton spin when the IPM hard gluon
coefficient is defined through Eq. (\ref{a48}) with $\Delta q^g$
defined as the quark distribution inside the gluon with
transverse squared momentum less than the factorization scale.
As a consequence, this anomalous gluonic
contribution in the IPM is free of infrared ambiguities.
We showed that if we accept the commonly used assumption of an
infinitely big momentum transfer, there is a $c$ quark
contribution to the spin in addition to the $u$, $d$ and $s$
quark contributions. The contribution from the massive quarks
is dependent on the factorization scale at which the polarized gluon
distribution is calculated. The $c$ quark contribution is small only
in the region $\mu^2 < 1\; GeV^2$, in which case the $s$ quark
contribution is also strongly affected.

We also calculated what would be the possible
anomalous corrections when the momentum transfer is in the region
of the present experiments. To perform such a calculation we
have to keep all terms in $m_q^2 /Q^2$ and $P^2 /Q^2$ in the
full photon-gluon cross section when calculating the hard
gluon coefficient. This means that we are including higher
twist effects and, although we use the complete result at
order $\alpha_s (Q^2)$, possible corrections coming from higher
order terms in $\alpha_s (Q^2)$ could be important and so our
calculation is incomplete. Even so, we think that our results
are more consistent than simply using the approximate expression (\ref{a431})
for the hard gluon coefficient in the case of massive quarks and relatively
low $Q^2$. The corrections due to finite $Q^2$ are
not small and we think they should be taken into account
when calculating the amount of spin carried by gluons.
When studying the $x$ dependence of the anomalous contribution,
we conclude that both 3 and 4 flavors give approximately the
same contribution inside the experimental region. But the amount
of polarized glue needed to fit the data is much smaller when
charm is included. Moreover, we showed that from the conceptual point of
view, it would be wrong not to include a fourth flavor.
\newline
\newline
\newline
We would like to thank to S. Bass, S. J. Brodsky, W. Melnitchouk and
G. Piller for helpful discussions. This work was supported
by the Australian Research Council and by CAPES (Brazil).

\addcontentsline{toc}{chapter}{\protect\numberline{}{References}}

\end{document}